%% file: main.tex
\documentclass{article}
\pdfoutput=1
\usepackage{graphicx}
\usepackage{hyperref}
\hypersetup{colorlinks,allcolors=black}
\usepackage[backend=biber, style=numeric,]{biblatex}
\usepackage{amsmath}
\usepackage{pgfplots}
\pgfplotsset{width=10cm}
\pgfplotsset{compat=1.18} 
\usepgfplotslibrary{external}
\title{M5: A Whole Genome Bacterial Encoder at Single Nucleotide Resolution}
\author{Agust Egilsson\footnote{ECAII, agust@ecaii.org}}
\date{June 2024}
\begin{document}

\maketitle
\begin{abstract}
A linear attention mechanism is described to extend the context length of an encoder only transformer, called M5 in this report, to a multi-million single nucleotide resolution foundation model pretrained on bacterial whole genomes. The linear attention mechanism used approximates a full quadratic attention mechanism tightly and has a simple and lightweight implementation for the use case when the key-query embedding dimensionality is low. The M5-small model is entirely trained and tested on one A100 GPU with 40gb of memory up to 196K nucleotides during training and 2M nucleotides during testing. We test the performance of the M5-small model and record notable improvements in performance as whole genome bacterial sequence lengths are increased as well as demonstrating the stability of the full multi-head attention approximation used as sequence length is increased.
\end{abstract}

\section{Introduction}
\input{introduction.tex}
\section{Background}
\input{background.tex}
\section{Methods}
\input{methods.tex}

\section{Results}
\input{results.tex}

\section{Discussion and Future Work}
\input{discussion.tex}
\printbibliography
\end{document}

%% file: introduction.tex
LLMs that incorporate DNA or RNA sequences need to successfully combine two factors: the extreme length of DNA sequences and the simplicity of the 4 main tokens involved. The most common way to combine these two factors is to use tokens representing k-mers to combine multiple nucleotides into a single token. Alternative approaches use single nucleotide resolution, i.e., k=1, as is the case with M5. The approach reported on here includes adjusting the architecture of the LL model to facilitate long sequences, directly taking advantages of the small token set and letting the model itself learn and optimize position encodings.

DNA and RNA segments, phased DNA and DNA in whole genomes represent a challenge for LLMs due to LLMs limitations on input context length and due to sub-optimal recall observed in many non-quadratic attention approaches, see \cite{jelassi2024repeat}. Recently, benchmarks have been introduced for comparing the performance of the various LLMs and models applied to DNA and RNA sequences, see \cite{marin2024bend}, \cite{kao2024advancing}, reflecting and providing an overview of some of the current approaches of measuring the performance of LLMs when specialized to DNA/RNA and applied to ultra long sequences.

For bacteria the genome segments are relatively short, ranging up to approximately 14 million nucleotides with most bacterial genomes below 5 million nucleotides, see \cite{diCenzo2017TheEvolution} and \cite{Martinez-Gutierrez2022GenomeScales}. M5 is an LLM designed to work with genomes in this size range at a single nucleotide resolution. It explores using a high number of attention heads, small key-query dimensionality, a family of approximations to the exponential function to linearize the attention mechanism and learned neighborhood based position embeddings. We hypothesize that this is an efficient way to work with DNA and RNA sequences at a single nucleotide resolution.

For testing we use M5-small, a linear encoder, progressively pretrained using masking of WGS segments of up to 196,608 nucleotides sampled from bacterial genomes. We report performance of testing the M5-small network on input segments of up to 2 million nucleotides using a single A100-40gb GPU both for the training and testing.

%% file: background.tex
Multiple papers reference using polynomials and kernel methods to linearize the softmax operators used in multi-head attention, BASED \cite{arora2024simple} uses a 2nd order Taylor series to approximate the exponential function; \cite{Katharopoulos2020TransformersAttention} by Angelos Katharopoulos et al. discusses polynomial linearization as well as using other kernel functions; Performers \cite{Choromanski2021RETHINKINGPERFORMERS} approximate attention using a method called ``Fast Attention Via positive Orthogonal Random features approach'' obtaining linear space and time complexity; symmetric and asymmetric kernels are discussed in \cite{Tsai2019TransformerKernel} by Yao Hung Hubert Tsai et al. The ReZero paper \cite{Bachlechner2021ReZeroDepth} uses zero initialized gated residual networks to remove reliance on layer normalization and improve training of extremely deep neural networks. The \cite{zhang2024hedgehog} paper by Michael Zhang et al. discusses using shallow trained feed-forward networks to match softmax attention weights resulting in linear attention networks that preserve the spikiness and monotonic properties of the softmax operator. 

In the paper \cite{alman2023fast} Josh Alman and Zhao Song provide theoretical arguments and proofs, including using the prerequisite of connecting the key-query dimension $d_k$ to the input length $N$, i.e., number of nucleotides. Inspired by the paper, and as a future direction of research, we consider the relationship $d_k \sim \log(N)$ in our full M5 report. 

Long context length models in biology that use single nucleotide resolution include EVO \cite{Nguyen2024SequenceEvo} at 131Kb at the time of the publication and HyenaDNA at 1 million nucleotides, \cite{nguyen2023hyenadna}. HyenaDNA swaps out attention for the "Hyena operator" and EVO uses the "StripedHyena" architecture. Both of the models are pre-trained using the next token prediction strategy, i.e., autoregressively.

%% file: methods.tex
\subsection{Linear attention used by M5}
\label{linearization-logic}
We outline the linear attention mechanism used by the M5 transformer encoder. Starting with a given key-query embedding dimension $d = d_k$ we approximate the value $\exp(q \circ k + m)$, where $\circ$ denotes the dot product of two vectors $q,k\in \mathbf{R}^{d_k}$ and $m$ is a number possibly depending on the $q$ and $k$ vector domains, using nonlinear transformations $\theta_m, \phi$ such that
\[\exp(q \circ k + m) \approx \theta_m(q)\circ\phi(k).\] This approximation of $\exp(q \circ k + m)$ can be made as exact as needed at the expense of additional compute on a given bounded interval using the fact that polynomials may be used to uniformly approximate the exponential function on a given bounded interval. The motivation of including the value $m$ in the formula is to direct the theoretical input $q \circ k + m$ into a bounded from below interval, with almost all of the values falling into a given bounded region where the approximation holds well.

First let's start with any given polynomial approximation, \[\exp(x) \approx \sum_{i=0}^{n}a_i x^i.\] If $x = q\circ k + m$ then rewriting this polynomial approximation is straightforward using the binomial formula

\begin{equation}
\label{binomial}
\sum_{i=0}^{n}a_i (q\circ k + m)^i = \sum_{i=0}^{n} \sum_{j=0}^i \binom{i}{j} a_i m^{i-j}(q \circ k)^j.
\end{equation} 
Using $(q \circ k)^j = (\sum_{l=1}^d q_l k_l)^j$ we obtain the formulation
\begin{equation}
\label{direct-sum}
(\sum_{l=1}^d q_l k_l)^j = \sum_{l_1, \ldots, l_j} q_{l_1} \cdots q_{l_j} k_{l_1} \cdots k_{l_j} = \oplus_{l_1, \ldots, l_j} q_{l_1} \cdots q_{l_j} \circ \oplus_{l_1, \ldots, l_j} k_{l_1} \cdots k_{l_j}
\end{equation}
where we are using dot product to combine the direct sums in \eqref{direct-sum}. Combining \eqref{binomial} and \eqref{direct-sum} we obtain the expression
\begin{equation}
    \sum_{i=0}^{n}a_i (q\circ k + m)^i = \sum_{j=0}^n \oplus_{l_1, \ldots, l_j} c_m^j q_{l_1} \cdots q_{l_j} \circ \oplus_{l_1, \ldots, l_j} k_{l_1} \cdots k_{l_j}
\end{equation} with 
\begin{equation}
    c_m^j = \sum_{i=j}^n \binom{i}{j} a_i m^{i-j}
\end{equation}
This allows us to write \eqref{binomial} as $\theta_m(q) \circ \phi(k)$, i.e., 
\begin{equation}
\label{theta-phi}
\exp(q\circ k + m) \approx \theta_m(q) \circ \phi(k)
\end{equation} 
with 
\begin{equation}
\theta_m(x) = c_m^0\oplus_{i_1} c_m^1 x_{i_1} \oplus_{i_1, i_2} c_m^2 x_{i_1}x_{i_2} \oplus\cdots\oplus_{i_1,\ldots, i_n} c_m^n x_{i_1} \cdots x_{i_n}
\end{equation} and
\begin{equation}
\phi(x) = 1\oplus_{i_1} x_{i_1} \oplus_{i_1, i_2} x_{i_1}x_{i_2} \oplus\cdots\oplus_{i_1,\ldots, i_n} x_{i_1} \cdots x_{i_n}
\end{equation} 
Instead of starting with a polynomial approximation of $\exp(x)$ we could equivalently start with an approximation of $\exp(x/\sqrt{d_k})$ since the softmax is normally scaled using $\sqrt{d_k}$, e.g., in the groundbreaking ``Attention Is All You Need'' paper \cite{Vaswani2017AttentionNeed}.
So we assume, in the below, that we are using a polynomial approximation 
\begin{equation}
\label{polynomial-approximation}
    \exp(x/\sqrt{d_k}) \approx \sum_{i=0}^{n}a_i x^i
\end{equation}

The difficulty with $\theta_m$ and $\phi$ is that these map $\mathbf{R}^d$ to $\mathbf{R}^{1+d+d^2+\cdots+d^n}$, so the dimensionality grows fast with $n$ and $d$ and the maps quickly become infeasible in any practical setting for large values of $n$ or $d$. On the one hand, $d_k$ can be reduced by mapping linearly the query and key projections to a smaller dimensional space, as in \cite{arora2024simple}. On the other hand, we can try to let n be small, e.g., $3$, and the key-query dimension $d=d_k$ be small also, say less than 10 and use a relatively high number of ``heads'' in a multi-head attention encoder. For a very small vocabulary, such as DNA at nucleotide resolution, we evaluate if this is a reasonable approach or not.
\hspace{1cm}
\subsubsection{Example} 
\begin{figure}
\centering
\begin{tikzpicture}
\pgfplotsset{width=8cm}
\begin{axis}[legend style={draw=none}, legend cell align={left}]
\addplot[color=red, domain=-2:5]{exp(x/2)};
\addlegendentry{$\scriptstyle e^\frac{x}{2}$}
\addplot[color=blue, domain=-2:5]{1.0017636 + 0.49488056*x + 0.12190779*x^2 + 0.02954964*x^3};
\addlegendentry{$\scriptscriptstyle 1.001764 + 0.494881x + 0.121908x^2 + 0.029550x^3$}
\end{axis}
\end{tikzpicture}
\caption{Example approximation on the interval [-1,2]}
\label{fig:poly-approx}
\end{figure}
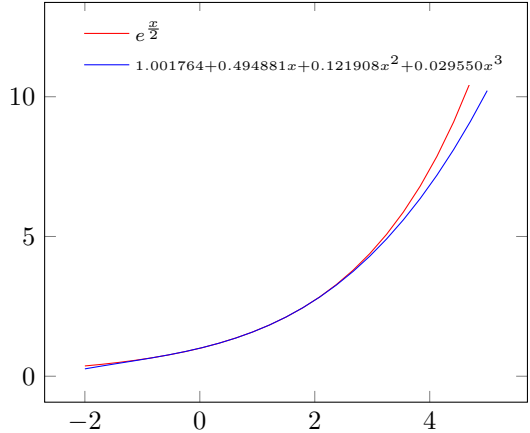

Empirically, if we are using $n=3, d_k=4$ and \[(a_0, a_1, a_2, a_3) = (1.0017636, 0.49488056, 0.12190779, 0.02954964)\] we obtain an excellent approximation to $\exp(x/\sqrt{d_k})$ on the interval $[-1,2]$, see Fig. \ref{fig:poly-approx}. The values of $(a_0, a_1, a_2, a_3)$ are chosen so that the integral of the squared difference between the polynomial and the scaled exponential function are minimized on the interval $[0,2]$, giving a total area under the squared difference of 9.5E-7\footnote{For this example, we are using the minimize function in SciPy \cite{Virtanen2020SciPyPython} and symbolically computing the integral using Wolfram Alpha PRO.} over the interval. 

Assuming we have a triad composed of query, key and value projections $Q\in\mathbf{R}^{N \times d_k}, K\in\mathbf{R}^{N \times d_k}$ and $V\in\mathbf{R}^{N \times d_v}$ we can, using the above, approximate a single row, $v_i'$, in the attention operator from \cite{Vaswani2017AttentionNeed} given as 
\begin{equation}
\label{attention_qkv}
    \text{Attention}(Q,K,V) = \text{softmax}(\frac{QK^t}{\sqrt{d_k}})V 
\end{equation}
using the formula
\begin{equation}
\label{linear-attention}
v_i' \approx \frac{\theta_m(q_i)^t \mathcal{K}_v}{\theta_m(q_i)^t \mathcal{K}}
\end{equation} with
\begin{equation}
    \mathcal{K}_v = \sum_{j=1}^N\phi(k_j)v_j^t \text{ and }
    \mathcal{K} = \sum_{j=1}^N\phi(k_j)
\end{equation} here we are using the vector to matrix convention $\mathbf{R}^n = \mathbf{R}^{n\times 1}$ to translate from dot products to matrix operations. Note that approximation \eqref{linear-attention} can be made arbitrarily precise on a given closed interval. The softmax function is invariant under constant shifts in the input logits, i.e., if $f_i(x_1, \ldots, x_n) = e^{x_i}/\sum_j e^{x_j}$ then $f_i(x_1+m, \ldots, x_n+m) = f_i(x_1, \ldots, x_n)$. Since the softmax normalizes each row in the hypothetical attention matrix then we can pick $m=m_i$ to be given by different values for each input position $i$ in \eqref{linear-attention}. 
\newline 
Here we let 
\begin{equation}
\label{def-m}
    m_i = ||q_i||_2 \max_j(||k_j||_2)
\end{equation}
this formula guarantees that $m_i\geq q_i \circ k_j$ for all $j\in\{1,\ldots,N\}$ and hence this allows us to assume that $x=q\circ k + m$ in approximation \eqref{polynomial-approximation} is always non-negative. Note that we never calculate $q_i \circ k_j$ directly for all $i$ and $j$ since that would require $N^2$ compute. We can however, use the inequality $-m_i \leq q_i \circ k_j \leq m_i$ and monitor $m_i$ during training to map out if our assumptions, regarding the interval for where approximation \eqref{polynomial-approximation} holds, are well-founded. To summarize, we have defined a family of transformations $\theta_{m_i}$
\begin{equation}
    \theta_{m_i}: \mathbf{R}^d \rightarrow \mathbf{R}^\frac{d^{n+1} - 1}{d - 1}\text{ for } i \in \{1,\ldots,N\}\text{ and } d = d_k
\end{equation}
where $N$ is the encoder context length such that attention mechanism \eqref{attention_qkv} may be linearized up to arbitrary precision using
\begin{equation}
\label{approx-result}
\exp(\frac{q_i \circ k_j + m_i}{\sqrt{d_k}}) \approx \theta_{m_i}(q_i)\circ\phi(k_j) 
\end{equation}
and where approximation \eqref{approx-result}, for each i, is no worse than $\epsilon_i \geq 0$ with
\begin{equation}
\label{epsilon-estimate}
\epsilon_i = \max_{0 \leq x \leq 2 m_i} | \exp(\frac{x}{\sqrt{d_k}}) - \sum_{k=0}^n a_k x^k|
\end{equation} 

\subsection{Compute required by the M5 attention mechanism}
In this section we approximate the exponential function with a polynomial of degree $n$ and evaluate the compute required by the attention mechanism \eqref{linear-attention} over context length $N$. Calculating \eqref{linear-attention} for all $i=1,\ldots, N$ requires compute on the order of 
\begin{equation}
\sim d_v d_k^n N
\end{equation} while computing quadratic attention requires compute on the order of \[(d_k + d_v)N^2.\] Assuming that $d_k << d_v$ we therefore obtain a speedup ratio of 
\begin{equation}
\label{speedup}
\sim \frac{(d_v + d_k)N^2}{d_v d_k^n N} \sim \frac{N}{d_k^n}.
\end{equation} If we let $d_k = d_\text{model}/h$ where $h$ is the number of heads, then the compute speedup or slowdown translates to
\begin{equation}
    \sim \frac{h^n N}{d_\text{model}^n}.
\end{equation}
Accordingly, we focus on models with a large number of heads, moderate model dimension and small values for $d_k$ such as $2,3$ or $4$. Note that the relationship $d_k = d_\text{model}/h$ is not a requirement for attention to work and we may deviate from it and rely on \eqref{speedup} instead.

\subsection{Position embeddings}
Training is performed on continuous slices of bacterial genomes with a repeated fixed token between individual chromosomes or plasmids. A learned position embedding is added to each nucleotide representing the genome segment, numbered from 1 to 4, the idea being to teach the network that the bacterial genome is made up of individual separated chromosomes and plasmids. We use a multi-layer learned convolutional network with max pool activations to map the neighborhood of each nucleotide to a position embedding that is then added to the nucleotide embedding (A,C,G,T, unknown, separator or mask). This is because we think of a position in the genome as being almost uniquely determined by its neighboring DNA sequence, hence we train a network to learn the position of a center nucleotide from its neighborhood. We set the neighborhood length to be 1024 nucleotides. The artifacts (RNA and proteins) produced by the bacterial genome during transcription and translation interact in the cell environment so interaction between elements should to a large degree be determined by the localized DNA sequence of each element instead of the absolute or relative position of tokens within the sequence.

For a future addition to this report, we include, the ability to switch on rotary positional embedding in the M5 model setting, see \cite{su2023roformer} for a discussion of the relative positional rotary encoding. The rotary positional embeddings are for the purpose of the current first version of this report switched off.

\subsection{Network training}
\label{network-training}
M5 is a transformer encoder only. It is trained using masking, see \cite{Devlin2019BERT:Understanding}, 12\% of known tokens are masked during training, 3\% are included as-is into the prediction. In addition to this a large continuous segment of random length up to a maximum of 4096 nucleotides, capped at 15\% of the context length, is hidden during training by replacing it with the mask token. The context length is increased gradually starting with context length of only 1024 and then doubled between sessions. Each session initializes the Adam learning rate, adjusts it to a new batch size, when needed, and uses a cosine learning schedule with warm-up. Weight-decay is used only for the the key-query transformation weight matrices and the gradients are clipped individually at 0.05. The key-query attention matrices (weights) are regulated using a small $l_2$ weight decay factor in some of the training runs in order to guarantee that approximation \eqref{polynomial-approximation} holds sufficiently well.

\subsection{M5-small training data}
We collected bacterial whole genome data from GTDB (Genome Taxonomy DataBase), see \cite{Parks2022GTDB:Taxonomy}, release 09-RS220 from April 24th, 2024. After processing the genomes downloaded and after scanning for keywords like ``complete genome'' and ``complete sequence'' in a description field we obtained approximately 7,000 whole genomes that we randomly divided into 80\%-10\%-10\% train, eval and test split. The total number of genomes downloaded from GTDB is much larger than the $\sim~ 7,000$ satisfying our criteria. For the purpose of evaluating the M5-small model architecture and evaluating its performance we used only the complete genomes as indicated by the aforementioned keywords inserted by the submitters of the data.

\subsection{Network architecture}
The original encoder found in \cite{Devlin2019BERT:Understanding} and \cite{Vaswani2017AttentionNeed} is our staring point. As explained in section \ref{linearization-logic} the attention mechanism is linearized by approximating the exponential function with a family of asymmetric kernel function pairs - one function per input position/token. The position embeddings come from a simultaneously learned CNN network and the top classification layer is similarly a multi-layer CNN network that is allowed to see a range of neighboring tokens before making a prediction. Layer normalization is applied after positional encodings have been added and layer normalization is used as the last step of each multi-head attention repeat. Additional skip connections are used and all skip connections are initialized so that the residual signal starts out as zero, this is achieved by initializing one sub-layer within each residual block with zeros only and is inspired by the logic found here \cite{Bachlechner2021ReZeroDepth} used to reduce the reliance on layer normalization in both cases.

%% file: results.tex
\subsection{Small M5 model}

The M5-small encoder is trained starting with a context length of 1,024 and then the context length is increased (normally doubled) between sessions until it reaches 196,608 nucleotides during training. For the initial small setup we use a key-query dimension of 4, $d_k=4$ and linearize the attention mechanism with a polynomial of degree 3, see \ref{linearization-logic} for the details. We train and test the model using only one A100 40gb GPU. For testing we use bacterial whole genome sequence segments of length up to 2,000,000 nucleotides. Everything, both testing and training is done at a single nucleotide resolution using only the 4 tokens A,C,G,T plus 3 meta-tokens (unknown, masking and a token representing both the separation of chromosomes and plasmids). Batch size is reduced from 16 to 1 due to memory restrictions as context length is increased and the learning rate schedule is modified as batch size is decreased or increased.

\subsection{Improved predictions as context length increases}

\begin{figure}
\centering

\begin{tikzpicture}
\pgfplotsset{width=10cm}
\begin{semilogxaxis}[
    log basis x=2,
    xlabel={model(s) testing context length},
    ylabel={CE for trained model(s)},
    xmin=1024, xmax=2097152,
    ymin=0.8, ymax=1.25,
    xtick={1024, 2048, 4096, 8192, 16384, 32768, 65536, 131072, 262144, 524288, 1048576, 2097152},
    ytick={0.9, 1.0, 1.1},
    legend pos=north west,
    xmajorgrids=true,
    ymajorgrids=true,
    grid style=dashed,
    legend cell align={left},
]

\addplot[
    color=red,
    mark=square,
    ]
    coordinates {
    (1024,1.0282) (2048,1.0030) (4096,0.9877) (8192,0.9691) (16384,0.9602) (32768,0.9515) (65536,0.9493) (131072, 0.9423) (196608, 0.9361)
    };
\addlegendentry{\small A: $\max$(train segment) = test context length}

\addplot[
    color=green,
    mark=square,
    ]
    coordinates {
    (1024,0.1579/0.15) (2048,0.1575/0.15) (4096,0.1581/0.15) (8192,0.1583/0.15) (16384,0.1590/0.15) (32768,0.1582/0.15) (65536,0.1570/0.15) (131072,1.0385) (262144,1.0352)
    };
\addlegendentry{\small B: train segments = 1,024 nucleotides}

\addplot[
    color=blue,
    mark=square,
    ]
    coordinates {
    (1024,0.9622) (2048,0.9640) (4096,0.9501) (8192,0.9419) (16384,0.9400) (32768, 0.9389) (65536, 0.9446) (131072, 0.9405) (262144,0.9389) (524288, 0.9358) (1048576, 0.9279) (1572864, 0.9109) (1835008, 0.9051) (2000000, 0.9010)
    };
\addlegendentry{\small C: train segments $\leq$ 196,608 nucleotides}
    
\end{semilogxaxis}
\end{tikzpicture}
\caption{Performance (CE Cross-Entropy) as context length is varied}
\label{fig:loss-context-length}
\end{figure}
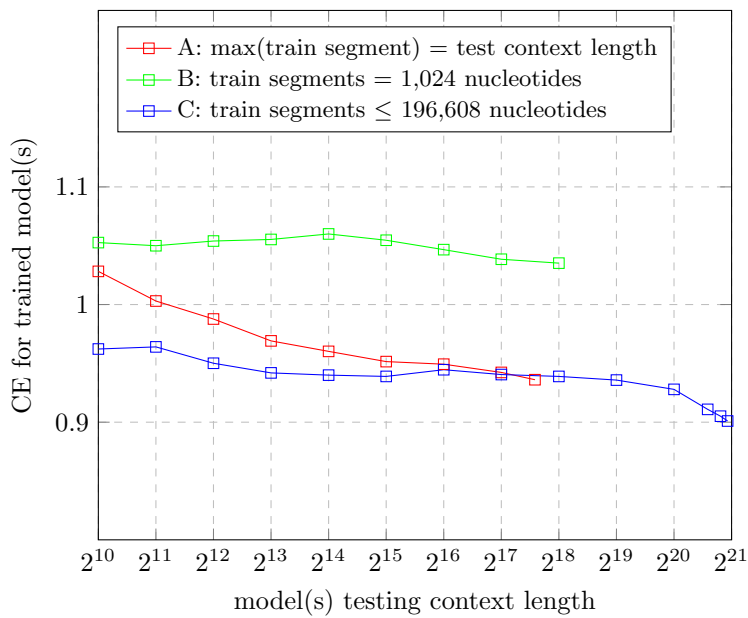

{\bf Figure \ref{fig:loss-context-length}} demonstrates, using cross-entropy measure, that there is a clear benefit of training on long sequences when it comes to inferencing about the bacterial genome. The figure shows the mean sample cross-entropy, on the y-axis, obtained using models trained on various sequence lengths, as detailed below, using the 10\% holdout test set. Cross-entropy is computed using the masked nucleotides, and measures how confidently the model on average is able to reconstruct the masked nucleotides, with zero representing complete confidence. As explained in section \ref{network-training} 12\% of the nucleotides are masked, 3\% are unchanged and included in the cross-entropy test compute. Additionally, during training a uniform random length continuous segment of up to a maximum of 15\% of the context length, capped at 4,096 nucleotides, is completely cleared of all nucleotide information. Plot {\bf A} (red) shows results from multiple models trained progressively on whole genome bacterial segments ranging from 1,024 to 196,608 nucleotides and then tested on segments equalling the longest training segment. Cross-entropy is reported for the model trained up to the given testing context input length shown on the x-axis. It is apparent from plot {\bf A} how the performance of our models improves steadily when trained on longer segments. Plot {\bf B} (green) shows the cross-entropy of a single model that is trained on nucleotide segments of length 1,024 only and then applied to longer DNA segments during the testing shown. Interestingly, when the context length is increased during inference the performance of this model does not appear to improve much if any. Plot {\bf C} (blue) shows the sample cross-entropy measured using a single model trained on, up to, bacterial whole genome sequence lengths of 196,608 nucleotides and then tested on various shorter and longer input context lengths shown on the x-axis. Even when this model is tested on the smallest input context length of 1,024 it by far outperforms the model trained explicitly only on that short sequence length. Combined, plots {\bf A, B} and {\bf C} show that training a model on long bacterial nucleotide sequences has clear benefits for predicting, using the M5-small model setup, when inferencing about shorter bacterial sequence segments as well as being beneficial for inferencing on longer sequences.

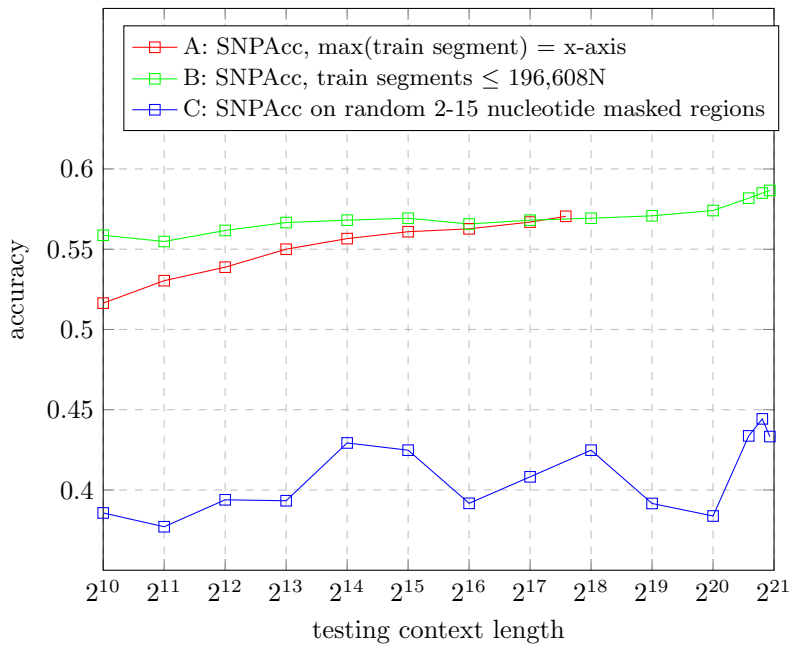
\begin{figure}
\centering

\begin{tikzpicture}
\pgfplotsset{width=10.5cm}
\begin{semilogxaxis}[
    log basis x=2,
    xlabel={testing context length},
    ylabel={accuracy},
    xmin=1024, xmax=2097152,
    ymin=0.35, ymax=0.70,
    xtick={1024, 2048, 4096, 8192, 16384, 32768, 65536, 131072, 262144, 524288, 1048576, 2097152},
    ytick={0.40, 0.45, 0.50, 0.55, 0.60},
    legend pos=north west,
    xmajorgrids=true,
    ymajorgrids=true,
    grid style=dashed,
    legend cell align={left},
]

\addplot[
    color=red,
    mark=square,
    ]
    coordinates {
    (1024, 0.5164) (2048, 0.5304) (4096, 0.5388) (8192, 0.5500) (16384, 0.5566) (32768, 0.5609) (65536, 0.5627) (131072, 0.5669) (196608, 0.5705)
    };
\addlegendentry{\small A: SNPAcc, $\max$(train segment) = x-axis}

\addplot[
    color=green,
    mark=square,
    ]
    coordinates {
    (1024, 0.5587) (2048, 0.5548) (4096, 0.5617) (8192, 0.5666) (16384, 0.5681) (32768, 0.5693) (65536, 0.5657) (131072, 0.5681) (262144, 0.5693) (524288, 0.5708) (1048576, 0.5741) (1572864, 0.5818) (1835008, 0.5850) (2000000, 0.5866)
    };
\addlegendentry{\small B: SNPAcc, train segments $\leq$ 196,608N}

\addplot[
    color=blue,
    mark=square,
    ]
    coordinates {
    (1024, 0.3857) (2048, 0.3771) (4096, 0.3939) (8192, 0.3933) (16384, 0.4293) (32768, 0.4248) (65536, 0.3917) (131072, 0.4082) (262144, 0.4248) (524288, 0.3916) (1048576, 0.3838) (1572864, 0.4337) (1835008, 0.4443) (2000000, 0.4333)
    };
\addlegendentry{\small C: SNPAcc on random 2-15 nucleotide masked regions}
    
\end{semilogxaxis}
\end{tikzpicture}
\caption{Single nucleotide prediction model accuracy (SNPAcc)}
\label{fig:accuracy-longest-model}
\end{figure}

{\bf Figure \ref{fig:accuracy-longest-model}} demonstrates, using prediction accuracy of masked nucleotides and short regions, the benefit of training on long sequences. The figure shows the accuracy of randomly masked (12\% masked + 3\% original) nucleotide predictions using the weights from models trained on bacterial segments of up to 196,608 bases. The accuracy is measured on the 10\% holdout set of the bacterial genomes collected as before. Plot {\bf A} (red) shows the single nucleotide model prediction accuracy achieved by models trained progressively up to and tested at the shown input context length (x-axis), i.e., 1,024 - 196,608 nucleotides. Plot {\bf B} (green) shows the accuracy (SNPAcc) for different input context length used during testing using a single model trained on sequences of up to length 196,608 nucleotides. Plot {\bf C} (blue) shows the the accuracy (SNPAcc) sampled on a randomly masked region of uniform random size between 2 and 15 nucleotides using the same encoder trained on sequences of up to length 196,608 nucleotides and tested for various sequence lengths as shown.

\subsubsection{Validity of the linear approximation to the softmax}
Here we look at the distribution of ``m'' values that are observed by the M5-small models. As demonstrated by formula \eqref{epsilon-estimate}, for our estimate of the approximation of the exponential function used to linearize the softmax in the M5 transformer encoder, we would like our approximation \eqref{polynomial-approximation} to hold in the interval $[0, 2 m_i]$. Inspired by eyeballing the approximations used in practice if we are using polynomials of odd degree such as the one seen in Figure \ref{fig:poly-approx} for $d_k=4$ we can shift the $m_i$ values by some small constant $\delta$, say -1 to increase the range where we have a solid approximation.

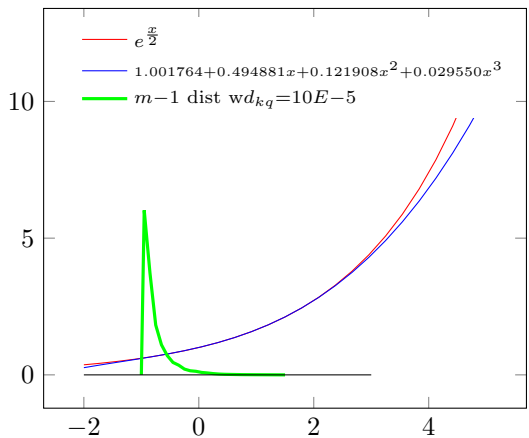
\begin{figure}
\centering
\begin{tikzpicture}
\def\multiplier{.15}

\pgfplotsset{width=8cm}
\begin{axis}[legend style={draw=none}, legend cell align={left}]

\addplot[color=red, domain=-2:5]{exp(x/2)};
\addlegendentry{$\scriptstyle e^\frac{x}{2}$}

\addplot[color=blue, domain=-2:5]{1.0017636 + 0.49488056*x + 0.12190779*x^2 + 0.02954964*x^3};
\addlegendentry{$\scriptscriptstyle 1.001764 + 0.494881x + 0.121908x^2 + 0.029550x^3$}

\addplot[
    color=green,
    mark=circle,
    domain=-2:5,
    very thick,
    ]
    coordinates {
(-1, 0.)
(-0.95,\multiplier*40.1196) 
(-0.85, \multiplier*24.8445)
(-0.75, \multiplier*12.1519) 
(-0.65, \multiplier*7.3329) 
(-0.55, \multiplier*4.7188) 
(-0.45, \multiplier*3.0351) 
(-0.35, \multiplier*2.3610)
(-0.25, \multiplier*1.4055) 
(-0.15, \multiplier*0.9993)
(-0.05, \multiplier*0.8670)
(0.05, \multiplier*0.5900)
(0.15, \multiplier*0.4668)
(0.25, \multiplier*0.3075)
(0.35, \multiplier*0.1973)
(0.45, \multiplier*0.1399)
(0.55, \multiplier*0.1087)
(0.65, \multiplier*0.0854)
(0.75, \multiplier*0.0600)
(0.85, \multiplier*0.0499)
(0.95, \multiplier*0.0400)
% (1.05, \multiplier*0.1295) - replace with for visualization, since this is the agreegate of all the remaining weights
(1.5, 0)
};
\addlegendentry{$\scriptstyle m - 1\text{ dist } {\text wd}_{kq} = 10E-5$}

\addplot[color=black, domain=-2:3]{0};

\end{axis}
\end{tikzpicture}
\caption{Distribution of $m-\delta$ for $d_k=4$}
\label{fig:mean-and-std-of-m}
\end{figure}

{\bf Figure \ref{fig:mean-and-std-of-m}} shows the distribution of the values $m_i - 1$ computed for all the attention heads, input tokens, and repeats for the 10\% test set and the M5-small encoder trained and tested using context length of 16,384 using formula \eqref{def-m}. For clarity we show the exponential function and the approximation used, demonstrating the tight approximation \eqref{polynomial-approximation} used to approximate the softmax on all applicable values in the test set. The distribution is impacted by the $l_2$ weight decay factor used to control the growth of the key and query weights (weight matrices), in this case it is set to 10E-5 for the key-query transformation matrices only. Empirically, we observe that $\text{dist}(m-\delta)$ is a one sided long tail distribution. The top of the distribution (40\%) is concentrated in the first histogram bucket (-1.00 to -0.90). Notably, the distribution implies a spiky dot product attention with 0.12\% of the $m_i$ entries larger than $2$ representing over 7 standard deviations from the mean, with mean for $m_i$ at 0.2223, and standard deviation of the distribution at 0.2558.

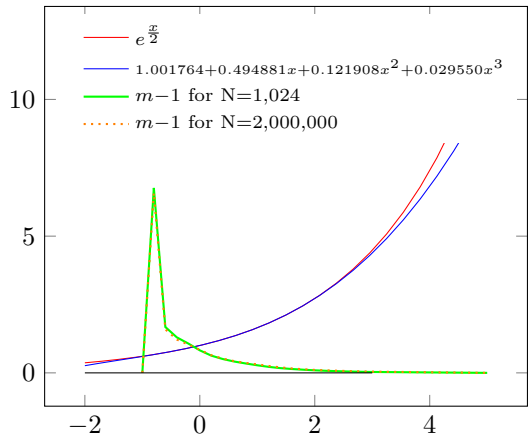
\begin{figure}
\centering
\begin{tikzpicture}
\def\multiplier{.15}

\pgfplotsset{width=8cm}
\begin{axis}[legend style={draw=none}, legend cell align={left}]

\addplot[color=red, domain=-2:5]{exp(x/2)};
\addlegendentry{$\scriptstyle e^\frac{x}{2}$}

\addplot[color=blue, domain=-2:5]{1.0017636 + 0.49488056*x + 0.12190779*x^2 + 0.02954964*x^3};
\addlegendentry{$\scriptscriptstyle 1.001764 + 0.494881x + 0.121908x^2 + 0.029550x^3$}

\addplot[
    color=green,
    mark=circle,
    domain=-2:5,
    thick,
    ] 
    coordinates {
(-1, 0.)
(-0.8,\multiplier*45.0930) 
(-0.6, \multiplier*11.1858)
(-0.4, \multiplier*8.6571) 
(-0.2, \multiplier*7.1494) 
(0.0, \multiplier*5.4931) 
(0.2, \multiplier*4.1999) 
(0.4, \multiplier*3.3729)
(0.6, \multiplier*2.7579) 
(0.8, \multiplier*2.2864)
(1.0, \multiplier*1.8504)
(1.2, \multiplier*1.4611)
(1.4, \multiplier*1.1582)
(1.6, \multiplier*0.9426)
(1.8, \multiplier*0.7703)
(2.0, \multiplier*0.6191)
(2.2, \multiplier*0.4991)
(2.4, \multiplier*0.4071)
(2.6, \multiplier*0.3297)
(2.8, \multiplier*0.2693)
(3.0, \multiplier*0.2191)
(5.0, 0)
};
\addlegendentry{$\scriptstyle m - 1\text{ for N=1,024}$}

\addplot[
    dotted,
    color=orange,
    mark=solid circle,
    domain=-2:5,
    thick,
    ] 
    coordinates {
(-1, 0.)
(-0.8,\multiplier*43.9066) 
(-0.6, \multiplier*10.6089)
(-0.4, \multiplier*8.1244 ) 
(-0.2, \multiplier*6.9050) 
(0.0, \multiplier*5.6172) 
(0.2, \multiplier*4.3141) 
(0.4, \multiplier*3.4522)
(0.6, \multiplier*2.8639) 
(0.8, \multiplier*2.4195)
(1.0, \multiplier*2.0607)
(1.2, \multiplier*1.6781)
(1.4, \multiplier*1.3407)
(1.6, \multiplier*1.0760)
(1.8, \multiplier*0.8726)
(2.0, \multiplier*0.7191)
(2.2, \multiplier*0.6107)
(2.4, \multiplier*0.5202)
(2.6, \multiplier*0.4317)
(2.8, \multiplier*0.3611)
(3.0, \multiplier*0.3000)
(5.0, 0)
};
\addlegendentry{$\scriptstyle m - 1\text{ for N=2,000,000}$}

\addplot[color=black, domain=-2:3]{0};

\end{axis}
\end{tikzpicture}
\caption{Distribution of $m-\delta$ for N=1,024 vs N=2,000,000}
\label{fig:m-small-N-vs-large-N}
\end{figure}

{\bf Figure \ref{fig:m-small-N-vs-large-N}} shows two distributions of the values $m_i - 1$ computed for all the attention heads, input tokens, and repeats for the model trained on sequences up to 196,608 nucleotides. One of the distributions is for the testing set sampled for segments of length N=1,024 and the other one is for the testing set sampled for segments of length N=2,000,000. Apparently, these two extremes in the context length only shift the distributions in very minor ways, even thought the computations of the $m_i$ values have to factor in the complete context input length as is evident by the use of the ``$\max$'' function over all indices in the input context from formula \eqref{def-m}.

%% file: discussion.tex
This report is a stepping stone in presenting a model that supports sufficient input context length to train on whole sequence bacterial genomes. We have presented the M5-small LLM foundation model supporting multi-million nucleotides as input during inference and efficiently progressively trained the models on segments, from the bacterial genome, of up to 196K tokens on a single 40gb GPU as well as supporting 2M tokens during inference on the same GPU. 

The model architecture may be optimized further to allow for larger input length on the same hardware. For example, the number of channels in the convolutional layers may be excessive and further testing will reveal the impact on performance of reducing the number of channels and at the same time the system could allocate more space required by further increasing the context length. 

As a future direction, we will report testing the full M5 model in a multi-GPU/TPU setting which allows us to scale the model context input. Further, testing the models ability to detect genome attributes such as transcription test sites and regulatory elements as well as measure fitness and conservation of regions is underway. Multiple architectural modification will be considered in future work such as investigating the benefits of the relationship $d_k \sim \log(N)$ between the key-query dimensionality and the input context length $N$.